\documentclass[letterpaper, 10 pt, conference]{ieeeconf}

\IEEEoverridecommandlockouts
\usepackage[T1]{fontenc}
\usepackage[utf8]{inputenc}
\usepackage{amsmath,amssymb}
\usepackage{graphicx}
\usepackage{booktabs}
\usepackage{makecell}
\usepackage{multirow}
\usepackage[table]{xcolor}
\usepackage{cite}
\graphicspath{{figures/}}

\definecolor{bestcell}{RGB}{246,198,205}
\definecolor{secondcell}{RGB}{211,228,252}
\definecolor{figlinkblue}{RGB}{0,92,175}
\definecolor{citelinkred}{RGB}{170,45,45}
\usepackage[colorlinks=true,linkcolor=figlinkblue,citecolor=citelinkred,urlcolor=figlinkblue]{hyperref}

\title{\LARGE \bf
WS-NeRF: A Mamba-Driven World-State-Aware Adaptive Deblurring\\
Neural Radiance Field
}

\author{Hang Jiang$^{1,2}$, Jinghao Wang$^{1,2}$, Yiming Zhang$^{1,2}$,\\
Xinhong Wang$^{1,2}$, Luwei Ran$^{1,2}$, and Yinfeng Yu$^{1,2,3,4,\dagger}$%
\thanks{$^{1}$School of Intelligent Science and Technology, Xinjiang University, Urumqi, China. Email: {\tt\small jianghang1aa@icloud.com}.}%
\thanks{$^{2}$School of Computer Science and Technology, Xinjiang University, Urumqi, China. Email: {\tt\small jianghang1aa@icloud.com}.}%
\thanks{$^{3}$Joint Research Laboratory for Embodied Intelligence, Xinjiang University, Urumqi, China. Email: {\tt\small yuyinfeng@xju.edu.cn}.}%
\thanks{$^{4}$Joint International Research Laboratory of Silk Road Multilingual Cognitive, Xinjiang University, Urumqi, China. Email: {\tt\small yuyinfeng@xju.edu.cn}.}%
\thanks{$^{\dagger}$Corresponding author: Yinfeng Yu ({\tt\small yuyinfeng@xju.edu.cn}).}%
}

\begin{document}

\maketitle
\thispagestyle{empty}
\pagestyle{empty}

\begin{abstract}
Neural Radiance Fields (NeRF) have attracted extensive attention in recent years due to their strong capability for high-quality 3D reconstruction and novel view synthesis from multi-view images. Existing methods usually rely on high-quality sharp inputs, while real-world image acquisition is highly susceptible to blur degradation, which severely affects the reconstruction quality of NeRF. In this paper, we propose a novel \emph{Mamba-driven world-state-aware adaptive deblurring neural radiance field}, termed \textbf{WS-NeRF}, to address image degradation and 3D inconsistency. We formulate the alternating optimization of radiance fields as a dynamic evolution process with temporal memory, and jointly exploit comprehensive multi-dimensional world states and a mixture-of-experts mechanism to dynamically adjust the confidence of deblurring priors. Experimental results show that WS-NeRF significantly improves blurry radiance field reconstruction quality, achieving better performance on PSNR, SSIM, and LPIPS, while exhibiting more stable iterative recovery behavior.
\end{abstract}

\section{INTRODUCTION}
\begin{figure*}[!t]
    \centering
    \includegraphics[width=0.8\textwidth]{"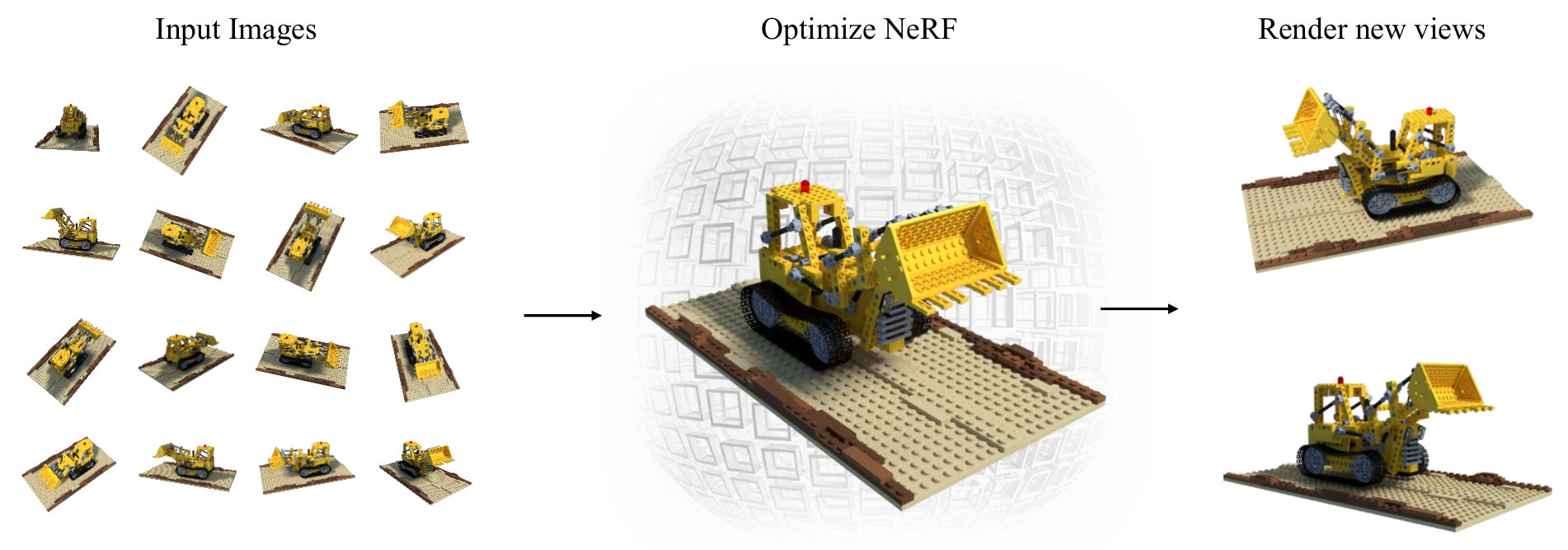"}
    \caption{Overview of the blur-aware radiance-field reconstruction task. Blurry input images are used to optimize the radiance field and synthesize restored novel views.}
    \label{fig:teaser_intro}
\end{figure*}

Since the seminal work of Mildenhall \emph{et al.}~\cite{mildenhall2020nerf}, Neural Radiance Fields (NeRF) have achieved photorealistic 3D representation and novel view synthesis by encoding scene geometry and appearance into multilayer perceptrons. Owing to their strong self-supervised learning capability and impressive rendering quality, NeRF-based methods have been widely applied in robot navigation, urban modeling, and virtual reality. However, the high-fidelity rendering ability of NeRF strongly depends on high-quality sharp inputs. In real scenarios, blur frequently occurs due to long exposure or camera motion, which seriously violates the physical assumptions of NeRF and leads to severe degradation of 3D reconstruction quality. Therefore, improving the 3D reconstruction capability of NeRF under blurred observations has become an important research topic. The blur-aware reconstruction task is illustrated in Fig.~\ref{fig:teaser_intro}.

Although many methods attempt to recover blurred radiance fields by alternating between 2D deblurring and 3D reconstruction, effectively coupling external image priors with 3D geometric constraints remains challenging. The intrinsic mismatch between single-view 2D restoration priors and global 3D spatial consistency introduces three major difficulties. First, there is a dimensional mismatch between 2D restoration priors and 3D geometric constraints. Second, prior fusion lacks a confidence-aware closed loop. Third, temporal modeling over optimization states is absent.

To address these challenges, we model radiance field deblurring as a dynamic evolution system with temporal memory. During training, we jointly leverage comprehensive multi-dimensional world states and a mixture-of-experts (MoE) mechanism to adaptively adjust the confidence of deblurring priors. Specifically, we design a world-state estimator that deeply fuses low-level spatial cues, high-frequency edge features, and deep semantic discrepancies extracted by a pretrained large vision model, DINOv2~\cite{oquab2023dinov2}, to diagnose spatial inconsistency in real time. The resulting state features are then fed into a Mamba-based temporal inference network~\cite{gu2023mamba}, endowing the system with long-range memory over optimization trajectories. Driven by both multi-dimensional state awareness and temporal memory, the adaptive controller collaborates with a MoE module to intelligently and dynamically allocate confidence to 2D restoration signals at each iteration. By balancing 2D prior restoration ability and 3D geometric constraints, the proposed method achieves both high-quality deblurring and stable scene convergence.

The main contributions of this paper are summarized as follows:
\begin{enumerate}
    \item We propose \textbf{WS-NeRF}, a novel adaptive state-guided deblurring framework that fundamentally resolves the inconsistency between introducing external 2D priors and maintaining 3D geometric constraints through dynamic perception and control.
    \item We construct a \textbf{multi-granularity world-state perception mechanism} that innovatively fuses spatial cues, restoration-aware signals, and deep semantic features from DINOv2 to sensitively monitor subtle geometric degradation during reconstruction.
    \item We design a \textbf{Mamba-based temporal module} for radiance field optimization trajectory modeling, enabling the system to exploit historical evolution information and effectively alleviate over-correction and oscillation commonly observed in alternating optimization.
\end{enumerate}

\section{RELATED WORK}
\subsection{Neural Radiance Fields and Novel View Synthesis}
NeRF implicitly models scene density and color with multilayer perceptrons, achieving remarkable success in novel view synthesis~\cite{mildenhall2020nerf}. Subsequent studies have further improved radiance-field representations in terms of rendering efficiency, representation capability, and view generalization.

\subsection{Radiance Field Restoration Under Blur}
Existing research on 3D reconstruction from blurry inputs can generally be divided into two categories. The first category is based on explicit degradation modeling. Methods such as Deblur-NeRF~\cite{ma2022deblurnerf}, DP-NeRF~\cite{lee2023dpnerf}, and BAD-NeRF~\cite{wang2023badnerf} simulate the physical degradation process by jointly optimizing blur kernels or continuous camera trajectories. Although these methods are physically interpretable, they rely heavily on parameterized assumptions, making it difficult to accurately fit the complex and non-uniform blur encountered in real-world scenarios. In addition, their training process is extremely time-consuming.

The second category is based on alternating optimization and prior guidance. To overcome the limitations of explicit modeling, methods such as PDRF~\cite{peng2023pdrf}, ExBluRF~\cite{lee2023exblurf}, DeepDeblurRF~\cite{choi2025deepdeblurrf}, and AFDU-NeRF~\cite{cao2025afdu} introduce 2D image deblurring priors. Related Gaussian-splatting-based deblurring methods, including Deblurring-3DGS~\cite{lee2024deblurring3dgs} and BAGS~\cite{peng2024bags}, further explore explicit scene representations under blur. By alternately optimizing the 2D deblurring network and the 3D radiance field, this strategy effectively combines the strong single-image restoration capability of 2D networks with the multi-view geometric consistency inherent in 3D representations.
\subsection{Multimodal Perception and Dynamic Fusion}
Related advances in embodied multimodal perception demonstrate the benefits of complementary sensory cues, collaborative observations, and external knowledge for robust navigation~\cite{YinfengICLR2022saavn,yu2023measuring,yang2026beyond,yu2025dope,li2025audio,zhang2025advancing,zhang2025iterative,yu2025dynamic}. Dynamic gating and representation fusion have also been investigated for audio-visual source separation and incomplete multimodal understanding~\cite{yu2025dgfnet,wang2025modality}, while multi-domain detail enhancement and attention-based feature fusion improve robustness in remote-sensing and speech applications~\cite{fu2025fsdenet,mattursun2024bss,zhang2024nonlinear,cao2024vnet}. Although these studies address tasks different from radiance-field restoration, their common principle of reliability-aware fusion motivates our adaptive regulation of heterogeneous 2D restoration cues and 3D geometric evidence.
\subsection{Error Accumulation in Iterative Restoration and Temporal State Modeling}
Most existing alternating deblurring frameworks statically treat the reliability of rendered guidance. Blindly trusting unconverged 3D priors at early stages can easily cause serious error accumulation. On the other hand, although world models and state-space models have shown strong capability in representing latent system states and mitigating catastrophic forgetting in long-sequence tasks, their use in complex 2D--3D radiance field restoration remains largely unexplored. To address this gap, we are the first to formulate iterative radiance-field deblurring as a dynamic evolution system. We exploit a world model to implicitly perceive residual restoration states and introduce Mamba to propagate long-range states across iterations, thereby adaptively regulating the strength of radiance field priors and fundamentally blocking cumulative errors caused by unreliable guidance.

\section{METHOD}
\subsection{Overall Framework}
\begin{figure*}[!t]
    \centering
    \includegraphics[width=0.9\textwidth]{"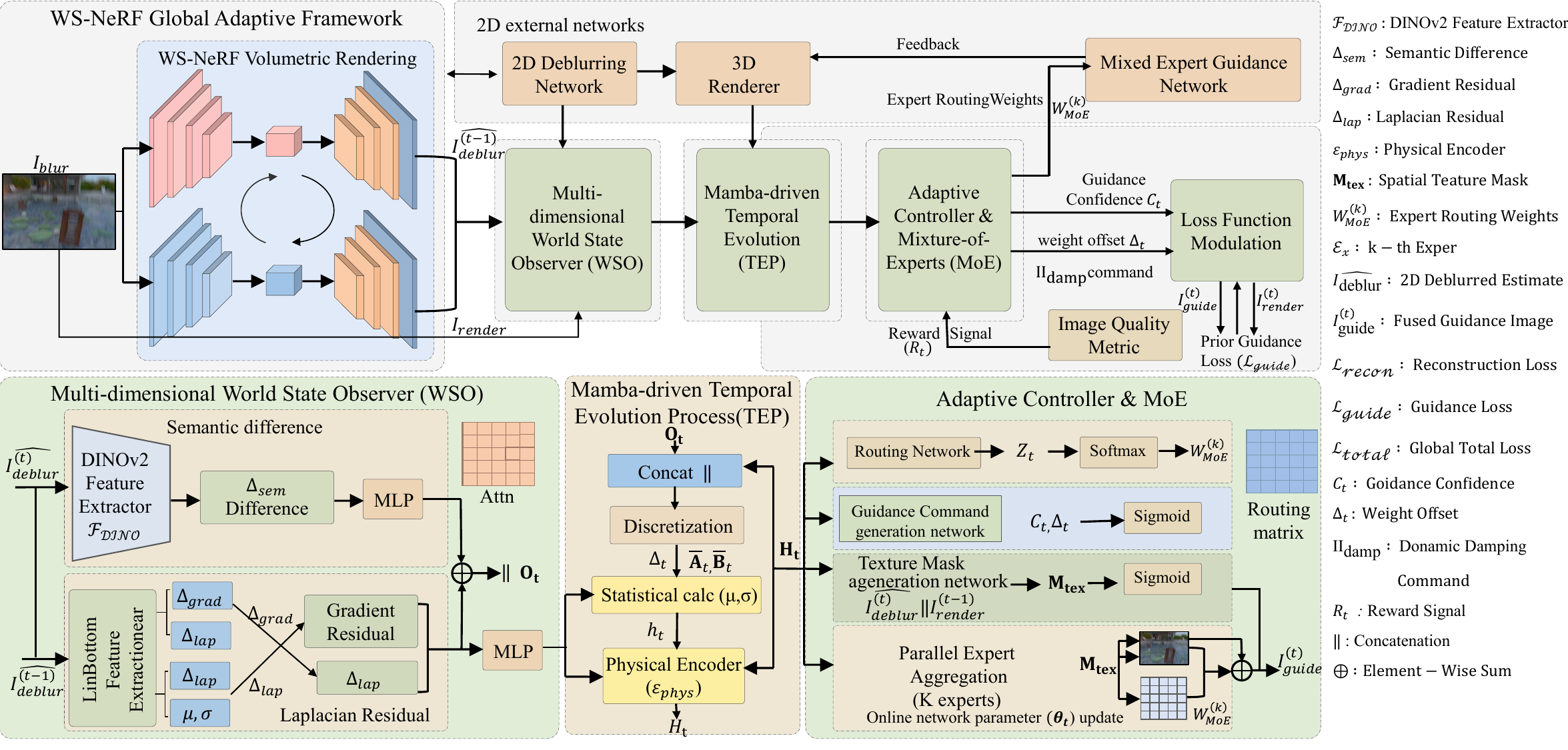"}
    \caption{Overall framework of WS-NeRF. The world-state observer, Mamba-driven temporal evolution process, and adaptive controller with mixture-of-experts jointly regulate prior guidance during radiance-field optimization.}
    \label{fig:framework}
\end{figure*}

This paper proposes WS-NeRF, an adaptive deblurring Neural Radiance Field framework driven by Mamba-based world state awareness. By leveraging a multi-dimensional world state estimator for error perception and Mamba for long-range temporal memory, the framework employs an adaptive controller to dynamically modulate the fusion ratio between 2D deblurring priors and 3D radiance fields, thereby achieving stable and synergistic optimization. The overall architecture is shown in Fig.~\ref{fig:framework}.
\subsection{Multi-Dimensional World-State Observer}
Traditional mean squared error is sensitive to high-frequency noise and cannot faithfully reflect the true quality of 3D reconstruction. Therefore, we build a multi-dimensional world-state observer that fuses high-level semantic cues and low-level physical features for state estimation. Acting as the sensing unit in the closed-loop control system, the observer innovatively combines deep semantic discrepancies extracted by a pretrained large vision model with low-level physical features derived from image gradients and Laplacian edges.

At the $t$-th iteration, the system receives the original blurry input $I_{\mathrm{blur}}$, the current 2D deblurred estimate $\hat{I}_{\mathrm{deblur}}^{(t)}$, and the rendered image from the 3D radiance field $I_{\mathrm{render}}^{(t)}$. Through a dual-stream architecture, the observer extracts the instantaneous state $O_t$ from both semantic and physical perspectives.

\subsubsection{Macroscopic Semantic and Microscopic Physical Dual-Stream Perception for State Estimation}
Macroscopic Semantic and Microscopic Physical Dual-Stream Perception 
To extract both global and local features, the estimator adopts a dual-stream architecture. Macroscopically, it leverages a pre-trained DINOv2 model to extract high-dimensional features from the deblurred estimate and the rendered image, computing their absolute error norm $\Delta_{sem}$. Being highly sensitive to global topological degradation, $\Delta_{sem}$ serves as an early warning to suppress 3D floaters and structural fragmentation caused by misleading 2D priors. Microscopically, Sobel and Laplacian operators are applied to extract first-order gradients and second-order high-frequency features, yielding low-level physical residuals $\Delta_{grad}$ and $\Delta_{lap}$. These residuals precisely localize texture distortions, perfectly compensating for the semantic features' lack of fine-grained, high-frequency awareness.

\subsubsection{State Aggregation and Observation Generation}
After extracting the above features, the system computes global spatial statistics (mean and standard deviation) of the residuals. The physical residuals are encoded by dedicated encoders and concatenated with the statistics to form a comprehensive low-level physical state:
\begin{equation}
\begin{aligned}
F_{\mathrm{phys}} = [&E_{\mathrm{grad}}(\Delta_{\mathrm{grad}}) \parallel \mu_{\mathrm{grad}} \parallel \sigma_{\mathrm{grad}} \parallel E_{\mathrm{lap}}(\Delta_{\mathrm{lap}}) \\
&\parallel \mu_{\mathrm{lap}} \parallel \sigma_{\mathrm{lap}}].
\end{aligned}
\end{equation}
Then, a multilayer perceptron fuses the physical state and the semantic discrepancy to generate the final instantaneous observation:
\begin{equation}
O_t = \mathrm{MLP}([\Delta_{\mathrm{sem}} \parallel F_{\mathrm{phys}}]) \in \mathbb{R}^{d}.
\end{equation}
This multi-dimensional observation escapes the limitations of single-direction pixel-level metrics.

\subsection{Mamba-Based Temporal Evolution and Dynamic Adjustment}
After obtaining the instantaneous observation $O_t$, directly making decisions based only on this signal would easily fall into the non-Markovian dilemma common in traditional alternating optimization. Therefore, WS-NeRF introduces a Mamba-based temporal propagation network as the state memory of the control system, enabling the framework to reference historical evolution trends and achieve temporally smooth, dynamically routed deblurring guidance.

\subsubsection{Selective State-Space Modeling and Gated Temporal Propagation}
The system treats the instantaneous observation $O_t$ as sequence input and updates the basic temporal hidden state $h_t$ via selective scanning based on a state-space model. To dynamically balance trust in current fluctuations and dependence on long-range memory, we further design an adaptive gated fusion mechanism on top of the Mamba output. Let $H_{t-1}$ denote the final hidden state from the previous iteration. The dynamic gate $g_t$ is computed as
\begin{equation}
g_t = \sigma(W_g[h_t, H_{t-1}] + b_g),
\end{equation}
and the current coherent state is updated by
\begin{equation}
H_t = g_t \odot \tanh(W_o h_t) + (1-g_t) \odot \tanh(W_h H_{t-1}).
\end{equation}
When the current observation contains burst noise caused by misleading 2D features, the gate automatically suppresses unreliable inputs and forces the system to rely more on the historical smooth state $H_{t-1}$.

\subsubsection{Dynamic Expert Weighting for Deblurring Prior Confidence}
The coherent state $H_t$ implicitly encodes the dynamic evolution of error convergence. It is further fed into a dynamic adjustment module to produce preference logits for different deblurring experts. A temperature-controlled softmax is then used to obtain the routing weights $W_{\mathrm{MoE}}^{(k)}$, which determine how aggressively the system should combine 3D priors and conservative 2D outputs at the current iteration.

\subsection{Progress-Aware Mixture-of-Experts Decision}
Based on $H_t$ and the routing weights, the adaptive controller outputs the final control commands.

\subsubsection{Pseudo-Supervised Online Update from Environment Feedback}
The system predicts the guidance confidence $C_t$ and a weight offset $\Delta_t$ from the coherent state. For online adaptation, the relative quality gain between consecutive rendered images is computed as an environment reward:
\begin{equation}
R_t = \mathrm{Metric}(I_{\mathrm{render}}^{(t)}) - \mathrm{Metric}(I_{\mathrm{render}}^{(t-1)}),
\end{equation}
where $\mathrm{Metric}(\cdot)$ denotes a no-reference image quality assessment function. When geometric corruption is detected ($R_t < 0$), the controller receives a penalty signal and rapidly decays the confidence to block the injection of erroneous priors.

\subsubsection{Mixture-of-Experts Guidance and Local Texture Fusion}
The controller maintains $K$ deblurring strategy experts $\mathcal{E}_k(\cdot)$ in parallel. To achieve spatially aware fine-grained control, a convolutional layer predicts a spatial texture mask $M_{\mathrm{tex}} \in [0,1]^{H \times W}$:
\begin{equation}
M_{\mathrm{tex}} = \mathrm{Sigmoid}(\mathrm{Conv}(\hat{I}_{\mathrm{deblur}}^{(t)} \parallel I_{\mathrm{render}}^{(t)})).
\end{equation}
The final fused guidance image is then constructed as
\begin{equation}
\begin{aligned}
I_{\mathrm{guide}}^{(t)} = \sum_{k=1}^{K} W_{\mathrm{MoE}}^{(k)}
\Big[&M_{\mathrm{tex}} \odot \mathcal{E}_k(I) \\
&+ (1-M_{\mathrm{tex}}) \odot \hat{I}_{\mathrm{deblur}}^{(t)} \Big].
\end{aligned}
\end{equation}
This mechanism allows the system to aggressively inject details in high-confidence regions while enforcing conservative fusion in risky edge areas.

\subsubsection{Progress-Aware Dynamic Damping Trigger}
To prevent over-correction during late-stage optimization, the controller includes a progress-aware safety guard. The reconstruction quality gain is smoothed with an exponential moving average over a temporal window, and a damping indicator $I_{\mathrm{damp}} \in \{0,1\}$ is defined as
\begin{equation}
I_{\mathrm{damp}}=
\begin{cases}
1, & \text{if } \bar{R}_t < \epsilon \text{ or } \mathrm{Var}(R_{t-W:t}) > \delta,\\
0, & \text{otherwise},
\end{cases}
\end{equation}
where $\epsilon$ is a near-zero reward threshold and $\delta$ is an oscillation tolerance. When damping is activated, the confidence output is overridden by an exponential decay and the routing weights are forced toward a conservative expert distribution.

\subsection{Joint Optimization Objective}

To achieve stable joint optimization of the radiance field and camera trajectories, WS-NeRF formulates a compound loss function modulated by a closed-loop mechanism:
\begin{equation}
\mathcal{L}_{total} = \mathcal{L}_{recon} + \alpha \cdot \left[ (1 - \mathbb{I}_{damp}) \cdot C_t \right] \cdot \mathcal{L}_{guide} + \gamma \mathcal{L}_{reg}
\end{equation}
\noindent\textit{Physical reconstruction loss} ($\mathcal{L}_{recon}$) constrains the mean squared error (MSE) between the synthesized blurry rendering and the real blurry input, serving as the fundamental photometric anchor.

\noindent\textit{Prior guidance loss} ($\mathcal{L}_{guide}$) leverages the fused image $I_{guide}^{(t)}$ to impose structural supervision, combining $L_1$ error and SSIM~\cite{wang2004ssim} on the intermediate sharp renderings to distill 2D high-frequency details into 3D space.

\noindent\textit{Closed-loop dynamic modulation} adjusts the prior weights in real time via the confidence score $C_t$ and the damping indicator $\mathbb{I}_{damp}$.

\section{EXPERIMENTS}
\subsection{Experimental Setup}
\subsubsection{Datasets}

\begin{figure}[!htbp]
    \centering
    \includegraphics[width=\columnwidth]{"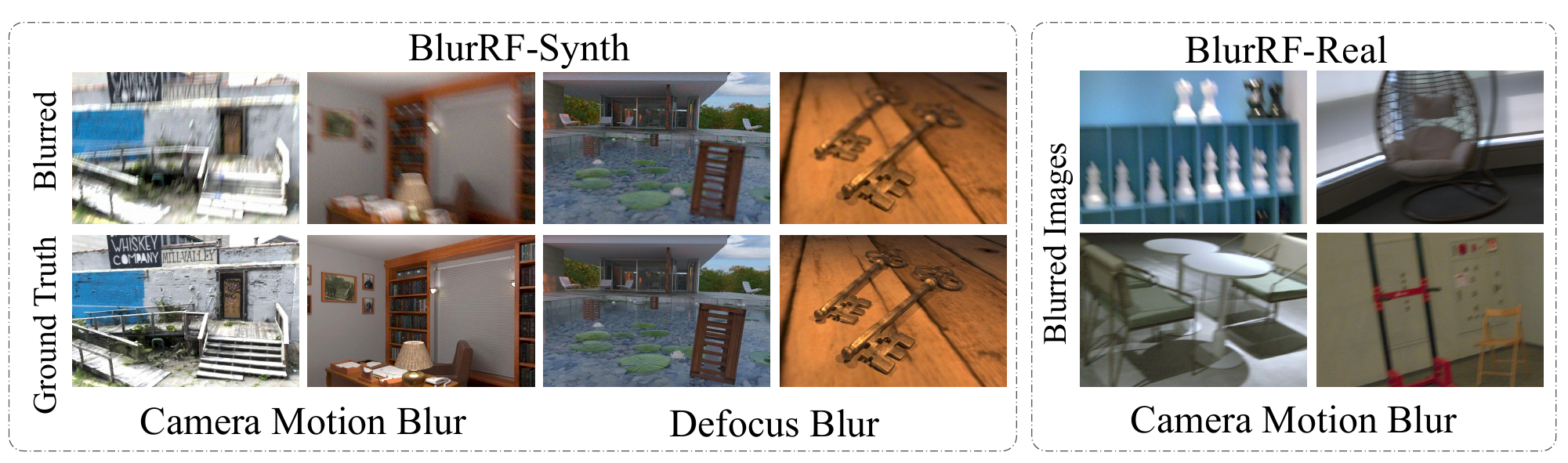"}
    \caption{Dataset examples used in the experiments, including synthetic motion blur, synthetic defocus blur, and real motion-blur scenes.}
    \label{fig:dataset}
\end{figure}

We adopt the benchmark dataset introduced by DeepDeblurRF~\cite{choi2025deepdeblurrf}. Specifically, (1) BlurRF-Synth contains 65 training scenes and 10 test scenes, covering both 6-DOF motion blur and defocus blur. In addition, physical noise is explicitly injected in the RAW space to better approximate real-world degradation. (2) BlurRF-Real contains five real low-light scenes with severe motion blur and high-frequency noise. Each scene provides 20--40 multi-view images for training, while 3--5 clear images are strictly reserved for novel view synthesis evaluation. Representative samples are shown in Fig.~\ref{fig:dataset}.
We evaluate synthetic scenes using PSNR, SSIM~\cite{wang2004ssim}, and LPIPS~\cite{zhang2018lpips}, and employ NIQE~\cite{mittal2013niqe} and ARNIQA~\cite{agnolucci2024arniqa} for blind quality assessment on real scenes. The full quantitative comparison and ablation summary are reported in Tables~\ref{tab:all_results} and~\ref{tab:ablation_summary}.
\subsubsection{Implementation Details}
WS-NeRF is implemented in PyTorch. To better support local feature extraction for the 2D deblurring network, we adopt image-patch-based ray sampling with a batch size of 4096 rays per iteration. The number of experts is set to $K=3$. The Mamba state dimension is set to $d=128$~\cite{gu2023mamba}. We use Adam for optimization, with the radiance field learning rate exponentially decayed from $5 \times 10^{-4}$ to $5 \times 10^{-5}$, and the adaptive controller learning rate fixed at $6 \times 10^{-4}$. Each scene is trained for 200k iterations on a single NVIDIA GPU. DINOv2 (ViT-Small) with frozen weights is used for high-level semantic perception~\cite{oquab2023dinov2}, NAFNet is adopted as the 2D deblurring executor, and the adaptive controller uses a dual-branch perception architecture with a shared 64-channel hidden layer and an additional Mamba state propagation module.
\begin{table*}[t]
\centering
\caption{Quantitative results of WS-NeRF and competing methods across all benchmarks. The best metric in each column is shown in bold.}
\label{tab:all_results}
\small
\setlength{\tabcolsep}{4.5pt}
\resizebox{\textwidth}{!}{%
\begin{tabular}{lccccccccccc}
\toprule
\multirow{2}{*}{Model} & \multicolumn{3}{c}{Camera Motion Blur} & \multicolumn{3}{c}{Defocus Blur} & \multicolumn{2}{c}{BlurRF-Real} & \multicolumn{3}{c}{BlurRF-SB} \\
\cmidrule(lr){2-4}\cmidrule(lr){5-7}\cmidrule(lr){8-9}\cmidrule(lr){10-12}
& PSNR ($\uparrow$) & SSIM ($\uparrow$) & LPIPS ($\downarrow$) & PSNR ($\uparrow$) & SSIM ($\uparrow$) & LPIPS ($\downarrow$) & NIQE ($\downarrow$) & ARNIQA ($\uparrow$) & PSNR ($\uparrow$) & SSIM ($\uparrow$) & LPIPS ($\downarrow$) \\
\midrule
Deblur-NeRF & 27.67 & 0.8340 & 0.1450 & 30.03 & 0.8727 & 0.1137 & 6.341 & 0.279 & 25.43 & 0.7264 & 0.2319 \\
BAD-NeRF & 21.74 & 0.5298 & 0.3969 & -- & -- & -- & 9.908 & 0.237 & 21.97 & 0.5231 & 0.4731 \\
DP-NeRF & 28.03 & 0.8412 & 0.1267 & 30.15 & 0.8763 & 0.0991 & 6.498 & 0.266 & 27.07 & 0.7940 & 0.1795 \\
\midrule
ExBluRF & 26.56 & 0.7823 & 0.1955 & -- & -- & -- & 7.479 & 0.279 & 24.16 & 0.6649 & 0.2659 \\
PDRF-10 & 28.33 & 0.8435 & 0.1495 & 30.03 & 0.8750 & 0.1225 & 6.243 & 0.266 & 26.66 & 0.7730 & 0.2124 \\
Deblurring-3DGS & 26.30 & 0.7729 & 0.1728 & 29.37 & 0.8545 & 0.1470 & 6.086 & 0.262 & 23.87 & 0.6459 & 0.2475 \\
BAGS & 27.41 & 0.8108 & 0.1382 & 29.90 & 0.8638 & 0.1152 & \textbf{5.837} & 0.290 & 24.49 & 0.6748 & 0.2391 \\
DeepDeblurRF & 29.93 & 0.8663 & 0.1163 & 32.49 & \textbf{0.9054} & 0.0976 & 6.086 & 0.262 & 29.06 & 0.8489 & 0.1516 \\
\textbf{WS-NeRF (Ours)} & \textbf{30.04} & \textbf{0.8699} & \textbf{0.1152} & \textbf{32.50} & 0.9046 & \textbf{0.0966} & 6.049 & \textbf{0.347} & \textbf{29.24} & \textbf{0.8519} & \textbf{0.1496} \\
\bottomrule
\end{tabular}
}
\end{table*}
\begin{table*}[t]
\centering
\caption{Ablation study of WS-NeRF. The best metric in each column is shown in bold.}
\label{tab:ablation_summary}
\small
\setlength{\tabcolsep}{5pt}
\resizebox{\textwidth}{!}{%
\begin{tabular}{lccccccccc}
\toprule
\multirow{2}{*}{Variant} & \multicolumn{3}{c}{Camera Motion Blur} & \multicolumn{3}{c}{Defocus Blur} & \multicolumn{3}{c}{BlurRF-SB} \\
\cmidrule(lr){2-4}\cmidrule(lr){5-7}\cmidrule(lr){8-10}
& PSNR ($\uparrow$) & SSIM ($\uparrow$) & LPIPS ($\downarrow$) & PSNR ($\uparrow$) & SSIM ($\uparrow$) & LPIPS ($\downarrow$) & PSNR ($\uparrow$) & SSIM ($\uparrow$) & LPIPS ($\downarrow$) \\
\midrule
w/o World State         & 29.84 & 0.8687 & 0.1164 & \textbf{32.55} & 0.9048 & 0.0967 & \textbf{29.26} & 0.8507 & 0.1498 \\
w/o Mamba               & 29.91 & 0.8672 & 0.1159 & 32.44 & 0.9051 & 0.0969 & 29.05 & 0.8502 & 0.1507 \\
w/o Multi-strategy Fusion & 29.93 & 0.8663 & 0.1163 & 32.49 & \textbf{0.9054} & 0.0976 & 29.06 & 0.8489 & 0.1516 \\
\textbf{Full Model}     & \textbf{30.04} & \textbf{0.8699} & \textbf{0.1152} & 32.50 & 0.9046 & \textbf{0.0966} & 29.24 & \textbf{0.8519} & \textbf{0.1496} \\
\bottomrule
\end{tabular}
}
\end{table*}
\subsection{Result Analysis}

Quantitatively, comparisons with existing mainstream baselines~\cite{ma2022deblurnerf,wang2023badnerf,lee2023dpnerf,lee2023exblurf,peng2023pdrf,lee2024deblurring3dgs,peng2024bags,choi2025deepdeblurrf} demonstrate that WS-NeRF achieves strong overall performance in handling complex blur degradation, benefiting from its state-aware perception and adaptive decision-making mechanism. As shown in Table~\ref{tab:all_results}, in motion blur scenarios, the proposed model attains an average PSNR of 30.04 and SSIM of 0.8699, consistently surpassing DeepDeblurRF. In defocus blur settings, WS-NeRF achieves superior LPIPS performance relative to the baseline methods. On the BlurRF-SB dataset, which involves more comprehensive degradation factors, our method exhibits a markedly better trade-off between perceptual quality and distortion than the competing approaches. Furthermore, under the real-world no-reference setting of BlurRF-Real, WS-NeRF demonstrates strong blind deblurring capability, achieving favorable performance in terms of NIQE.

\begin{figure}[!htbp]
    \centering
    \includegraphics[width=\columnwidth]{"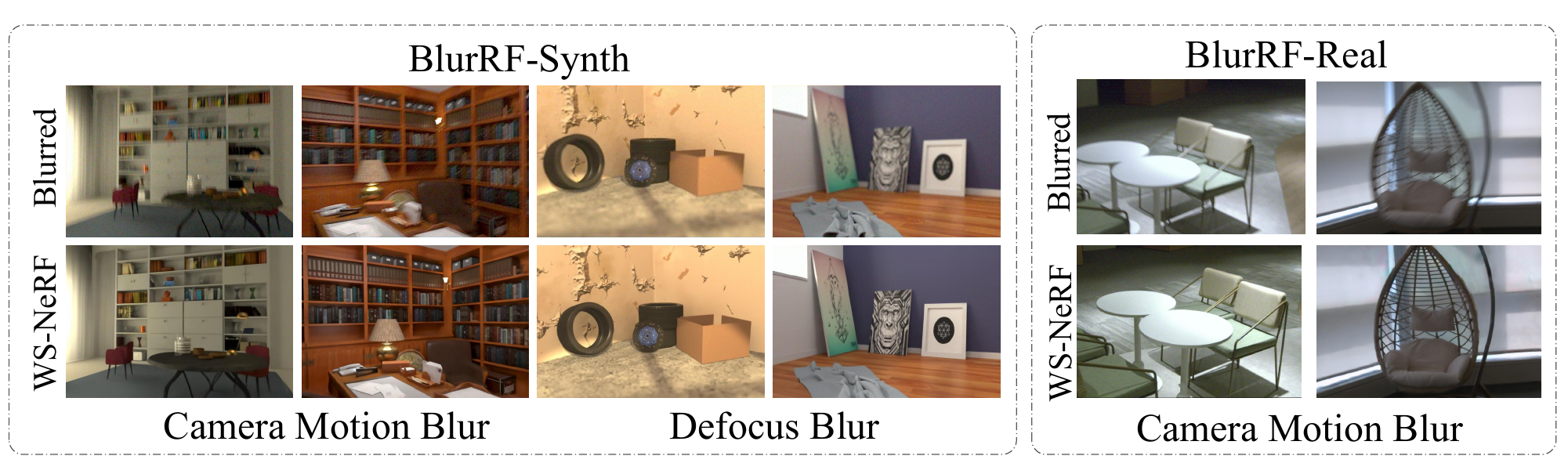"}
    \caption{Qualitative comparison on synthetic and real blurry radiance-field benchmarks. WS-NeRF restores sharper structures while preserving cross-view geometry.}
    \label{fig:qualitative_comparison}
\end{figure}

Qualitatively, under severe nonlinear blur degradation, WS-NeRF achieves finer detail restoration while better preserving three-dimensional geometric structures. Visualization results on the BlurRF synthetic and BlurRF real datasets are shown in Fig.~\ref{fig:qualitative_comparison}.

\subsection{Ablation Study}

\begin{figure}[!htbp]
    \centering
    \includegraphics[width=0.7\columnwidth]{"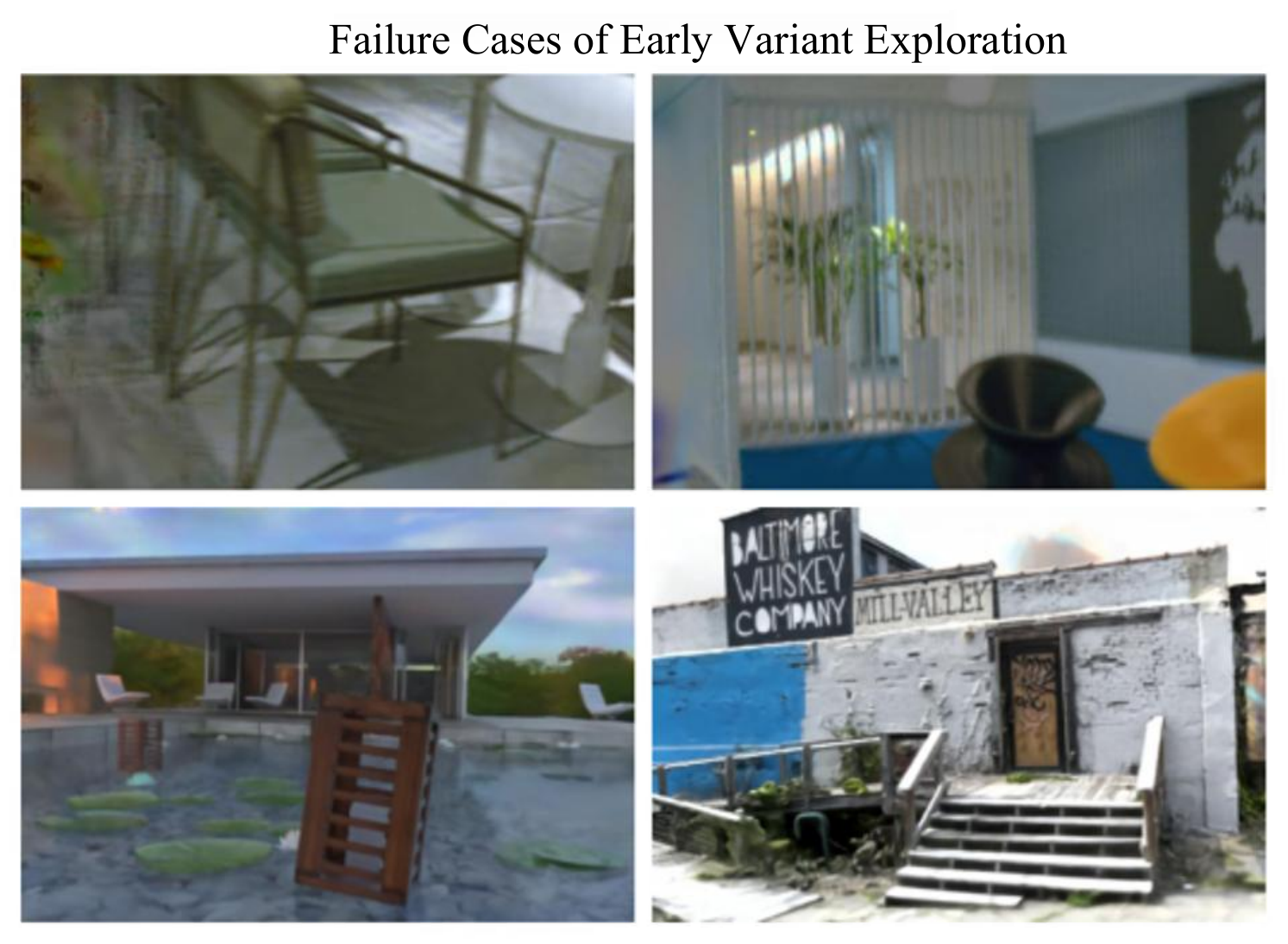"}
    \caption{Early-stage failure case in alternating radiance-field restoration. Unreliable rendered guidance can introduce artifacts and motivate state-aware adaptive control.}
    \label{fig:early_failure_case}
\end{figure}

We further validate the contributions of world-state perception (WS), the Mamba temporal propagation module, and the multi-strategy fusion mechanism. As shown in Table~\ref{tab:ablation_summary}, removing any component degrades performance on at least part of the benchmark suite. Among them, the multi-strategy fusion module contributes the largest single-module improvement, mainly because it dynamically balances the guidance ratio between 2D priors and 3D physical constraints. The Mamba module also yields a clear gain by establishing long-range temporal memory and reducing convergence oscillation. Fig.~\ref{fig:early_failure_case} further illustrates the early-stage artifacts caused by unreliable guidance, which motivates the adaptive damping design. Combined together, these modules form the complete WS-NeRF framework and deliver the most balanced overall performance.

\section{DISCUSSION AND CONCLUSION}
In this paper, we proposed WS-NeRF, an adaptive deblurring neural radiance field framework. By combining a multi-dimensional world-state evaluator with a Mamba-based temporal propagation network, the proposed method establishes a closed-loop dynamic routing mechanism for intelligent modulation of 2D deblurring priors and 3D physical constraints. The dynamic damping mechanism further suppresses systematic artifacts in alternating optimization. Extensive benchmark results indicate that WS-NeRF consistently improves reconstruction quality in extreme blur scenarios and significantly broadens the applicability of radiance fields in realistic degraded environments.

Limitations include sensitivity to initial hyperparameters and poor 2D prior generalization in textureless or underexposed scenes. Future work will investigate prior-free generative physical inversion and joint pose-radiance calibration under severe degradation for enhanced stability.

\section*{ACKNOWLEDGMENT}

This research was financially supported by Xinjiang University Training Program of Innovation and Entrepreneurship for Undergraduates (Grant No. 202510755141).


\bibliographystyle{ieeetr}
\bibliography{references}

@inproceedings{mildenhall2020nerf,
  author = {Mildenhall, Ben and Srinivasan, Pratul P. and Tancik, Matthew and Barron, Jonathan T. and Ramamoorthi, Ravi and Ng, Ren},
  title = {{NeRF}: Representing Scenes as Neural Radiance Fields for View Synthesis},
  booktitle = {Proceedings of the European Conference on Computer Vision},
  pages = {405--421},
  year = {2020}
}

@inproceedings{ma2022deblurnerf,
  author = {Ma, Li and Li, Xiaoyu and Liao, Jing and Zhang, Qi and Wang, Xuan and Wang, Jue and Sander, Pedro V.},
  title = {{Deblur-NeRF}: Neural Radiance Fields from Blurry Images},
  booktitle = {Proceedings of the IEEE/CVF Conference on Computer Vision and Pattern Recognition},
  pages = {12861--12870},
  year = {2022}
}

@inproceedings{wang2023badnerf,
  author = {Wang, Peng and Zhao, Lingzhe and Ma, Ruijie and Liu, Peidong},
  title = {{BAD-NeRF}: Bundle Adjusted Deblur Neural Radiance Fields},
  booktitle = {Proceedings of the IEEE/CVF Conference on Computer Vision and Pattern Recognition},
  pages = {4170--4179},
  year = {2023}
}

@inproceedings{lee2023dpnerf,
  author = {Lee, Dogyoon and Lee, Minhyeok and Shin, Chajin and Lee, Sangyoun},
  title = {{DP-NeRF}: Deblurred Neural Radiance Field with Physical Scene Priors},
  booktitle = {Proceedings of the IEEE/CVF Conference on Computer Vision and Pattern Recognition},
  pages = {12386--12396},
  year = {2023}
}

@inproceedings{peng2023pdrf,
  author = {Peng, Cheng and Chellappa, Rama},
  title = {{PDRF}: Progressively Deblurring Radiance Field for Fast Scene Reconstruction from Blurry Images},
  booktitle = {Proceedings of the AAAI Conference on Artificial Intelligence},
  volume = {37},
  pages = {2029--2037},
  year = {2023},
  doi = {10.1609/aaai.v37i2.25295}
}

@inproceedings{lee2023exblurf,
  author = {Lee, Dogyoon and Oh, Jeongtaek and Rim, Jaesung and Cho, Sunghyun and Lee, Kyoung Mu},
  title = {{ExBluRF}: Efficient Radiance Fields for Extreme Motion Blurred Images},
  booktitle = {Proceedings of the IEEE/CVF International Conference on Computer Vision},
  pages = {17639--17648},
  year = {2023}
}

@inproceedings{lee2024deblurring3dgs,
  author = {Lee, Byeonghyeon and Lee, Howoong and Sun, Xiangyu and Ali, Usman and Park, Eunbyung},
  title = {Deblurring {3D} Gaussian Splatting},
  booktitle = {Computer Vision -- ECCV 2024},
  pages = {127--143},
  year = {2025},
  doi = {10.1007/978-3-031-73636-0_8}
}

@inproceedings{peng2024bags,
  author = {Peng, Cheng and Tang, Yutao and Zhou, Yifan and Wang, Nengyu and Liu, Xijun and Li, Deming and Chellappa, Rama},
  title = {{BAGS}: Blur Agnostic Gaussian Splatting Through Multi-Scale Kernel Modeling},
  booktitle = {Computer Vision -- ECCV 2024},
  pages = {293--310},
  year = {2025},
  doi = {10.1007/978-3-031-72989-8_17}
}

@inproceedings{choi2025deepdeblurrf,
  author = {Choi, Haeyun and Yang, Heemin and Han, Janghyeok and Cho, Sunghyun},
  title = {Exploiting Deblurring Networks for Radiance Fields},
  booktitle = {Proceedings of the IEEE/CVF Conference on Computer Vision and Pattern Recognition},
  pages = {6012--6021},
  year = {2025}
}

@inproceedings{cao2025afdu,
  author = {Cao, Qinkang and Ma, Xiaolin and Wang, Chenyang and Kuang, Hailan and Liu, Xinhua},
  title = {{AFDU-NeRF}: Deblurring {NeRF} with Adaptive-Aware Fusion and Iterative Data Updating},
  booktitle = {Proceedings of the IEEE 6th International Seminar on Artificial Intelligence, Networking and Information Technology},
  pages = {493--496},
  year = {2025},
  doi = {10.1109/AINIT65432.2025.11035722}
}

@article{oquab2023dinov2,
  author = {Oquab, Maxime and others},
  title = {{DINOv2}: Learning Robust Visual Features without Supervision},
  journal = {arXiv preprint arXiv:2304.07193},
  year = {2023}
}

@article{gu2023mamba,
  author = {Gu, Albert and Dao, Tri},
  title = {Mamba: Linear-Time Sequence Modeling with Selective State Spaces},
  journal = {arXiv preprint arXiv:2312.00752},
  year = {2023}
}

@article{wang2004ssim,
  author = {Wang, Zhou and Bovik, Alan C. and Sheikh, Hamid R. and Simoncelli, Eero P.},
  title = {Image Quality Assessment: From Error Visibility to Structural Similarity},
  journal = {IEEE Transactions on Image Processing},
  volume = {13},
  number = {4},
  pages = {600--612},
  year = {2004}
}

@inproceedings{zhang2018lpips,
  author = {Zhang, Richard and Isola, Phillip and Efros, Alexei A. and Shechtman, Eli and Wang, Oliver},
  title = {The Unreasonable Effectiveness of Deep Features as a Perceptual Metric},
  booktitle = {Proceedings of the IEEE/CVF Conference on Computer Vision and Pattern Recognition},
  pages = {586--595},
  year = {2018}
}

@article{mittal2013niqe,
  author = {Mittal, Anish and Soundararajan, Rajiv and Bovik, Alan C.},
  title = {Making a Completely Blind Image Quality Analyzer},
  journal = {IEEE Signal Processing Letters},
  volume = {20},
  number = {3},
  pages = {209--212},
  year = {2013}
}

@inproceedings{agnolucci2024arniqa,
  author = {Agnolucci, Lorenzo and Galteri, Leonardo and Bertini, Marco and Del Bimbo, Alberto},
  title = {{ARNIQA}: Learning Distortion Manifold for Image Quality Assessment},
  booktitle = {Proceedings of the IEEE/CVF Winter Conference on Applications of Computer Vision},
  pages = {189--198},
  month = jan,
  year = {2024}
}

@inproceedings{YinfengICLR2022saavn,
  author = {Yu, Yinfeng and Huang, Wenbing and Sun, Fuchun and Chen, Changan and Wang, Yikai and Liu, Xiaohong},
  title = {Sound Adversarial Audio-Visual Navigation},
  booktitle = {The Tenth International Conference on Learning Representations},
  year = {2022}
}

@inproceedings{yu2023measuring,
  title = {Measuring Acoustics with Collaborative Multiple Agents},
  author = {Yu, Yinfeng and Chen, Changan and Cao, Lele and Yang, Fangkai and Sun, Fuchun},
  booktitle = {Proceedings of the Thirty-Second International Joint Conference on Artificial Intelligence},
  pages = {335--343},
  year = {2023}
}

@article{yang2026beyond,
  title = {Beyond Textual Knowledge: Leveraging Multimodal Knowledge Bases for Enhancing Vision-and-Language Navigation},
  author = {Yang, Dongsheng and Yu, Yinfeng and Wang, Liejun},
  journal = {Information Processing \& Management},
  volume = {63},
  number = {6},
  pages = {104766},
  year = {2026},
  publisher = {Elsevier}
}

@inproceedings{yu2025dope,
  title = {{DOPE}: Dual Object Perception-Enhancement Network for Vision-and-Language Navigation},
  author = {Yu, Yinfeng and Yang, Dongsheng},
  booktitle = {Proceedings of the 2025 International Conference on Multimedia Retrieval},
  pages = {1739--1748},
  year = {2025}
}

@inproceedings{yu2025dgfnet,
  title = {{DGFNet}: End-to-End Audio-Visual Source Separation Based on Dynamic Gating Fusion},
  author = {Yu, Yinfeng and Sun, Shiyu},
  booktitle = {Proceedings of the 2025 International Conference on Multimedia Retrieval},
  pages = {1730--1738},
  year = {2025}
}

@article{fu2025fsdenet,
  title = {{FSDENet}: A Frequency and Spatial Domains Based Detail Enhancement Network for Remote Sensing Semantic Segmentation},
  author = {Fu, Jiahao and Yu, Yinfeng and Wang, Liejun},
  journal = {IEEE Journal of Selected Topics in Applied Earth Observations and Remote Sensing},
  year = {2025},
  publisher = {IEEE}
}

@inproceedings{mattursun2024bss,
  title = {{BSS-CFFMA}: Cross-Domain Feature Fusion and Multi-Attention Speech Enhancement Network Based on Self-Supervised Embedding},
  author = {Mattursun, Alimjan and Wang, Liejun and Yu, Yinfeng},
  booktitle = {2024 IEEE International Conference on Systems, Man, and Cybernetics},
  pages = {3589--3594},
  year = {2024},
  organization = {IEEE}
}

@article{zhang2024nonlinear,
  title = {Nonlinear Regularization Decoding Method for Speech Recognition},
  author = {Zhang, Jiang and Wang, Liejun and Yu, Yinfeng and Xu, Miaomiao},
  journal = {Sensors},
  volume = {24},
  number = {12},
  pages = {3846},
  year = {2024},
  publisher = {MDPI}
}

@inproceedings{li2025audio,
  title = {Audio-Guided Dynamic Modality Fusion with Stereo-Aware Attention for Audio-Visual Navigation},
  author = {Li, Jia and Yu, Yinfeng and Wang, Liejun and Sun, Fuchun and Zheng, Wendong},
  booktitle = {International Conference on Neural Information Processing},
  pages = {346--359},
  year = {2025},
  organization = {Springer}
}

@inproceedings{zhang2025advancing,
  title = {Advancing Audio-Visual Navigation Through Multi-Agent Collaboration in 3D Environments},
  author = {Zhang, Hailong and Yu, Yinfeng and Wang, Liejun and Sun, Fuchun and Zheng, Wendong},
  booktitle = {International Conference on Neural Information Processing},
  pages = {502--516},
  year = {2025},
  organization = {Springer}
}

@article{zhang2025iterative,
  title = {Iterative Residual Cross-Attention Mechanism: An Integrated Approach for Audio-Visual Navigation Tasks},
  author = {Zhang, Hailong and Yu, Yinfeng and Wang, Liejun and Sun, Fuchun and Zheng, Wendong},
  journal = {arXiv preprint arXiv:2509.25652},
  year = {2025}
}

@article{yu2025dynamic,
  title = {Dynamic Multi-Target Fusion for Efficient Audio-Visual Navigation},
  author = {Yu, Yinfeng and Zhang, Hailong and Zhu, Meiling},
  journal = {arXiv preprint arXiv:2509.21377},
  year = {2025}
}

@inproceedings{wang2025modality,
  title = {Modality-Invariant Bidirectional Temporal Representation Distillation Network for Missing Multimodal Sentiment Analysis},
  author = {Wang, Xincheng and Wang, Liejun and Yu, Yinfeng and Jiao, Xinxin},
  booktitle = {ICASSP 2025--2025 IEEE International Conference on Acoustics, Speech and Signal Processing},
  pages = {1--5},
  year = {2025},
  organization = {IEEE}
}

@inproceedings{cao2024vnet,
  title = {{VNet}: A {GAN}-Based Multi-Tier Discriminator Network for Speech Synthesis Vocoders},
  author = {Cao, Yubing and Li, Yongming and Wang, Liejun and Yu, Yinfeng},
  booktitle = {2024 IEEE International Conference on Systems, Man, and Cybernetics},
  pages = {4384--4389},
  year = {2024},
  organization = {IEEE}
}

\end{document}